\documentclass[twocolumn]{aastex631}

\pdfoutput=1

\usepackage{newtxtext,newtxmath}
\usepackage[T1]{fontenc}

\usepackage{graphicx}	
\usepackage{amsmath}	
\usepackage{bm}         
\usepackage{ulem}       
\usepackage{hyperref}

\newcommand{\C}{\texttt{C}} 
\newcommand{\python}{\texttt{Python}}
\newcommand{\jupyter}{\textsc{Jupyter}}
\newcommand{\reb}{\textsc{rebound}}
\newcommand{\rebx}{\textsc{rebound\small{x}}}
\newcommand{\mercury}{\textsc{mercury}}
\newcommand{\swift}{\textsc{swift}}
\newcommand{\sse}{\textsc{sse}}

\newcommand{\Nbody}{\textit{N}-body}

\graphicspath{ {./images/} }

\shorttitle{The Yarkovsky effect in REBOUNDx}
\shortauthors{Ferich et al.}

\begin{document}

\title{The Yarkovsky effect in REBOUNDx}

\author{Noah Ferich}
\affiliation{Department of Astrophysical \& Planetary Sciences, University of Colorado Boulder, Boulder, CO 80309, USA \\}

\author{Stanley A. Baronett}
\affiliation{Department of Physics and Astronomy, University of Nevada, Las Vegas, Box 454002, 4505 S. Maryland Pkwy., Las Vegas, NV 89154-4002, USA\\}
\affiliation{Nevada Center for Astrophysics, University of Nevada, Las Vegas, Box 454002, 4505 S. Maryland Pkwy., Las Vegas, NV 89154-4002, USA\\}

\author{Daniel Tamayo}
\affiliation{Department of Astrophysical Sciences, Princeton University, 4 Ivy Ln, Princeton, NJ 08544, USA}

\author{Jason H. Steffen}
\affiliation{Department of Physics and Astronomy, University of Nevada, Las Vegas, Box 454002, 4505 S. Maryland Pkwy., Las Vegas, NV 89154-4002, USA\\}
\affiliation{Nevada Center for Astrophysics, University of Nevada, Las Vegas, Box 454002, 4505 S. Maryland Pkwy., Las Vegas, NV 89154-4002, USA\\}

\correspondingauthor{Noah Ferich}
\email{noah.ferich@colorado.edu}



\begin{abstract}

To more thoroughly study the effects of radiative forces on the orbits of small, astronomical bodies, we introduce the Yarkovsky effect into \rebx, an extensional library for the \Nbody\ integrator \reb. Two different versions of the Yarkovsky effect (the “Full Version” and the “Simple Version”) are available for use, depending on the needs of the user. We provide demonstrations for both versions of the effect and compare their computational efficiency with another previously implemented radiative force. In addition, we show how this effect can be used in tandem with other features in \rebx\ by simulating the orbits of asteroids during the asymptotic giant branch phase of a 2 $M_{\sun}$ star. This effect is made freely available for use with the latest release of \rebx.

\end{abstract}


\section{Introduction} 
\label{sec:intro}

Radiation pressure \citep{Nichols1903}, Poynting-Robertson drag \citep{Poynting1904, Robertson1937}, the Yarkovsky effect \citep{Radzievskii1954, Peterson1976}, and the YORP (Yarkovsky-O’Keefe-Radzievskii-Paddack) effect \citep{Radzievskii1954, Paddack1969, OKeefe1976} are four radiative forces that can be important for the orbital evolution of small bodies.  Both radiation pressure and Poynting-Robertson drag (PR drag) arise from the absorption and scattering of radiation by dust and other particles \citep{Burns1979}.  Radiation pressure is generally a force that propels particles away from their host stars, while PR drag is a force that causes particles to lose momentum and spiral inwards \citep{Veras2015}. 

Unlike radiation pressure and PR drag, the Yarkovsky and YORP effects appear only when considering objects' angular momentum \citep{Vokrouhlicky2015}.  Both phenomena occur after a radiation-induced temperature gradient forms on the surface of a body, causing an uneven emission of radiation between its hemispheres. As an object rotates relative to the substellar point, this imbalance in the emission creates a net force on the object.  While both effects are based on this principle, the Yarkovsky effect changes the orbit of a body, while the YORP effect creates a torque that changes the body's spin. In addition, the YORP effect only operates on non-spherical, asymmetric objects while the Yarkovsky effect can operate on spherical ones.  Even with the differences between them, all of these effects can have important consequences on the dynamical evolution of small bodies; this motivates developing tools to numerically model these phenomena.

\reb\footnote{Documentation is available at \url{https://rebound.readthedocs.io}.} is an open-source, \Nbody\ integrator for gravitational dynamics \citep{Rein2012}.  Flexible and efficient, \reb\ can be used in \C\ and is also available as a \python\ package. \rebx\footnote{Documentation is available at \url{https://reboundx.readthedocs.io}.} is an extensional library to \reb\ that contains additional effects and forces (general relativity, tidal dissipation, etc.) that users can add to simulations \citep{Tamayo2020}.  This library includes radiative forces \citep{Tamayo2020}, but contained only radiation pressure and PR drag and was lacking both the Yarkovsky and YORP effects.  This paper introduces the Yarkovsky effect as another radiative force available in \rebx\ and shows its capabilities and potential for future work.

We present two separate versions of the Yarkovsky effect available for use in \rebx: the Full Version and the Simple Version.  While both of these versions are based on the Yarkovsky effect model derived in \citet{Veras2015}, the Simple Version includes modifications done by \citet{Veras2019} that compromise long-term accuracy for a reduction in computational time and the number of necessary parameters.  More detail will be given in later sections on the differences between these two versions and when to appropriately use them. Please note that rotational perturbations created by the YORP effect are not included within this \rebx\ module.  YORP requires detailed models of the surface of a body \citep{Nesvorny2007, Scheeres2008} that are beyond the scope of this paper.  For now, we assume that all bodies are spherical and have a uniform density.  Adding the YORP effect into \rebx\ will be left for future work. 

The Yarkovsky effect has previously been added to other \Nbody\ integrators, including \mercury\ \citep{Chambers1999} and \swift\ \citep{Duncan1997}. While these integrators have their own advantages, \reb\ contains features that make it better suited for a wide variety of studies. \reb\ contains a larger array of different integrating schemes, providing nine different options for numerical integrators while \swift\ and \mercury\ only provide four and five, respectively. In addition, \reb\ comes standard with features such as collision tracking and the ability to create simulation archives that save particle and orbital parameters at certain times during an ongoing simulation. Features like these would have to be manually added to these other codes by users. \reb\ will also always yield results that are identical across all platforms and compilers. Finally, \rebx\ gives users greater flexibility when running simulations in \reb, allowing them to easily add multiple effects at once to simulations (see Section \ref{sec:combining_effects}).

Section~\ref{sec:yark_in_rebx} describes the main equations that govern the Yarkovsky effect in \rebx. In Section~\ref{sec:full_version}, we introduce the Full Version of the effect and demonstrate one of its possible applications, and in Section~\ref{sec:simple_version}, we introduce the Simple Version and show how an analytic equation describing the evolution of an object's semi-major axis can be derived from it under specific circumstances. Finally, we show how the Yarkovsky effect can be combined with other previously implemented \rebx\ effects and test how they perform together in Section~\ref{sec:combining_effects}.

\section{The Yarkovsky Effect in REBOUNDx}
\label{sec:yark_in_rebx}

Equation (27) from \citet{Veras2015} provides the model for calculating the acceleration on a body due to the Yarkovsky effect.  It is implemented by both versions of the Yarkovsky effect in \rebx.
\begin{equation}
    \left(\frac{d\textit{\textbf{v}}}{dt}\right)= \frac{3kL(1-\alpha)}{16\pi \rho Rcr^2}\mathbb{Y}\textit{\textbf{i}},
    \label{eq:veras2015}
\end{equation}
where $L$ is the luminosity of the object's host star\footnote{$L$ can be constant or time-dependent.  Section~\ref{sec:yark_effect_parameter_interpolation} gives an example of how the luminosity can be configured to change over time in a simulation.} and $c$ is the speed of light.  In addition, $\alpha$, $R$, and $\rho$ are the bond albedo, radius, and density of the body, and $r$ is its distance from the star.  The 3 $\times$ 3 matrix $\mathbb{Y}$ is the rotation matrix that contains the physics of the Yarkovsky effect, and $\textit{\textbf{i}}$ is the relativistically-corrected direction of the incoming radiation.

The parameter $k$ is a constant between 0 and 0.25; it was created to avoid the need to model the complex spin behavior of most asteroid-like objects.  Realistically, its value will change over time, but in these simulations, it remains constant to reduce complexity.  If the target’s rotation speed approaches the critical rotation speed at which the target will begin to break up \citep{Walsh2008, Vokrouhlicky2015}, then $k \rightarrow 0$.  If the target’s rotation period approaches its orbital period, then $k \rightarrow 0.25$.

Equation~(\ref{eq:veras2015}) includes the physics for both the diurnal and seasonal components of the Yarkovsky effect.  Both of these contributions are based on the same mechanism: the heated side of a body will release photons that are both more abundant and more energetic than those from the dark side. This leads to a net force on the body opposite the emission direction \citep{Bottke2002}.  The changes in the orbital parameters of an object depend on how this heated side is oriented when this radiation is emitted.  The diurnal contribution comes from the temperature differences created in the body due to its rotation with respect to its orbit.  This component can cause an increase or decrease in the semi-major axis of a body, depending on the orientation of its spin axis.  The seasonal contribution comes from temperature differences created due to the specific angular momentum of an object in its orbit.  This component will always cause a decrease in the semi-major axis of the body and does not depend on the orientation or magnitude of the object's spin.  Figure 1 from \citet{Bottke2002} provides a visual diagram of these two contributions, and Section~\ref{sec:full_version_demonstration} provides more information on how they compare and depend on different model parameters.  Both of these components are required for a complete model of the Yarkovsky effect.

We will now discuss the equations that comprise both $\mathbb{Y}$ and $\textit{\textbf{i}}$. The vector $\textit{\textbf{i}}$ is described by the following:
\begin{equation}
    \textit{\textbf{i}} = \left(1-\frac{\textit{\textbf{v}} \cdot \textit{\textbf{r}}}{cr}\right)\frac{\textit{\textbf{r}}}{r}-\frac{\textit{\textbf{v}}}{c},
\end{equation}
where $\textit{\textbf{r}}$ is the position vector of the body and $\textit{\textbf{v}}$ is its velocity vector.  If $\lvert \lvert \textit{\textbf{v}} \rvert \rvert << c $ then $\textit{\textbf{i}} \approx \textit{\textbf{r}} / r$, which is the unit vector in the direction of the incoming radiation.  The rotation matrix $\mathbb{Y}$ is the product of two matrices:
\begin{equation}
    \mathbb{Y} = \mathbb{R}_{Y}(\textit{\textbf{s}}, \phi)\mathbb{R}_{Y}(\textit{\textbf{h}}, \xi),
\end{equation}
where $\mathbb{R}_{Y}(\textit{\textbf{s}}, \phi)$ is the contribution from the diurnal component of the Yarkovsky effect and $\mathbb{R}_{Y}(\textit{\textbf{h}}, \xi)$ is the contribution from the seasonal component of the effect.  The vector $\textit{\textbf{s}}$ is the spin axis of the body, $\phi$ is the thermal lag angle along the equator of the body, $\textit{\textbf{h}}$ is the object's specific angular momentum axis, and $\xi$ is the thermal lag angle in the orbital plane of the body.  The equations for $\phi$ and $\xi$ are based off the specific 1D conduction model found in \citet{Broz2006}:
\begin{equation}
    \tan(\phi) = \left(1+\frac{1}{2}\left(\frac{\sigma \epsilon}{\pi ^ 5}\right)^\frac{1}{4}\left(\frac{\Sigma^\frac{1}{2}}{\Gamma}\right)\left(\frac{L(1-\alpha)}{r^2}\right)^\frac{3}{4}\right)^{-1},
    \label{eq:phi}
\end{equation}
\begin{equation}
    \tan(\xi) = \left(1+\frac{1}{2}\left(\frac{\sigma \epsilon}{\pi ^ 5}\right)^\frac{1}{4}\left(\frac{\Pi^\frac{1}{2}}{\Gamma}\right)\left(\frac{L(1-\alpha)}{r^2}\right)^\frac{3}{4}\right)^{-1},
    \label{eq:xi}
\end{equation}
where $\sigma$ is the Stefan-Boltzmann constant, $\epsilon$ is the object's emissivity, $\Sigma$ is its rotational period, $\Pi$ is its orbital period, and $\Gamma$ is its thermal inertia.  Both $\phi$ and $\xi$ only have meaningful values between $0^\circ$ and $45^\circ$. 

The full equations and derivations for $\mathbb{R}_{Y}(\textit{\textbf{s}}, \phi)$ and $\mathbb{R}_{Y}(\textit{\textbf{h}}, \xi)$ can be found in \citet{Veras2015}.  Section~\ref{sec:full_version} and Section~\ref{sec:simple_version} will describe how the two versions of the Yarkovsky effect are implemented in \rebx.

\section{Full Version}
\label{sec:full_version}

The Full Version of the Yarkovsky effect in \rebx\ is directly based on the model for the Yarkovsky effect found in \citet{Veras2015}.  Setting a \rebx\ parameter called "ye\_flag" to 0 on bodies in \reb\ simulations will apply this version of the effect to these chosen targets.  Users must input all of the necessary parameters contained in the model for this effect to work.  As is standard for all effects in \rebx, the parameters must be inputted with the same units as the \reb\ simulation where the effects are being added.  

The large number of parameters included in this version of the effect provides a high level of detail for the perturbations created by the Yarkovsky effect on a particular object.  However, this means long simulations containing many bodies with the Full Version active will require a large amount of computational time (see Section~\ref{sec:time_performance}).

\subsection{Demonstration}
\label{sec:full_version_demonstration}

To demonstrate the Full Version of the effect, we perform a simple parameter study that explores how the obliquity, radius, and thermal inertia of an asteroid affect the strength of the Yarkovsky effect.  This is similar to a study \citet{Carruba2017} performed to create Figure 15 in their study of the Veritas asteroid family.  

We create a 0.01 Myr-long simulation that contains the Sun and an asteroid in a circular, non-inclined orbit at a distance of 3.165802 au, the semi-major axis of 7231 Veritas \citep{Carruba2017}.  This value was chosen to place the asteroid at a realistic distance within the Veritas asteroid family.  Table~\ref{tab:parameter_table} lists the unchanging physical properties of the asteroid. The density, rotation period, and albedo are the same values that \citet{Carruba2017} used in their creation of Figure 15.  Because a value for the emissivity was not stated in \citet{Carruba2017}, we have chosen to use an estimate for the emissivity of 101955 Bennu from \citet{Emory2014}. Given that 101955 Bennu and the members of the Veritas family are all carbonaceous asteroids and generally similar in composition, the emissivity of 101955 Bennu should be a reasonable estimate for the emissivity of a Veritasian asteroid.

Each time the simulation is run, the asteroid is given a new combination for its obliquity, radius, and thermal inertia.  The object can have radii values of 10 m, 100 m, and 1000 m; obliquity values of $0^\circ$, $30^\circ$, $60^\circ$, $90^\circ$, $120^\circ$, $150^\circ$, and $180^\circ$; and a distribution of 50 thermal inertia values ranging from 0.1 to 4000.  The change in semi-major axis of the asteroid is recorded at the end of each simulation. 

\begin{table}
    \centering
    \caption{Unchanging Physical Properties of the Asteroid in the Full Version Parameter Study}
    \begin{tabular}{lcc}
    \hline
    \hline
         Property & Value and Uncertainty & Reference\\
         \hline
         Density & $1300$ \footnote{in units of kg m$^{-3}$.} & \citet{Carruba2017}\\
         Rotation Period & \textbf{$6$} \footnote{in units of hours.} & \citet{Carruba2017}\\
         Emissivity & $0.90 \pm 0.05\ $ & \citet{Emory2014}\\
         Albedo & $0.07$ & \citet{Carruba2017}\\
         \hline
    \end{tabular}
    \label{tab:parameter_table}
\end{table}

\begin{figure}
	\includegraphics[width=\columnwidth]{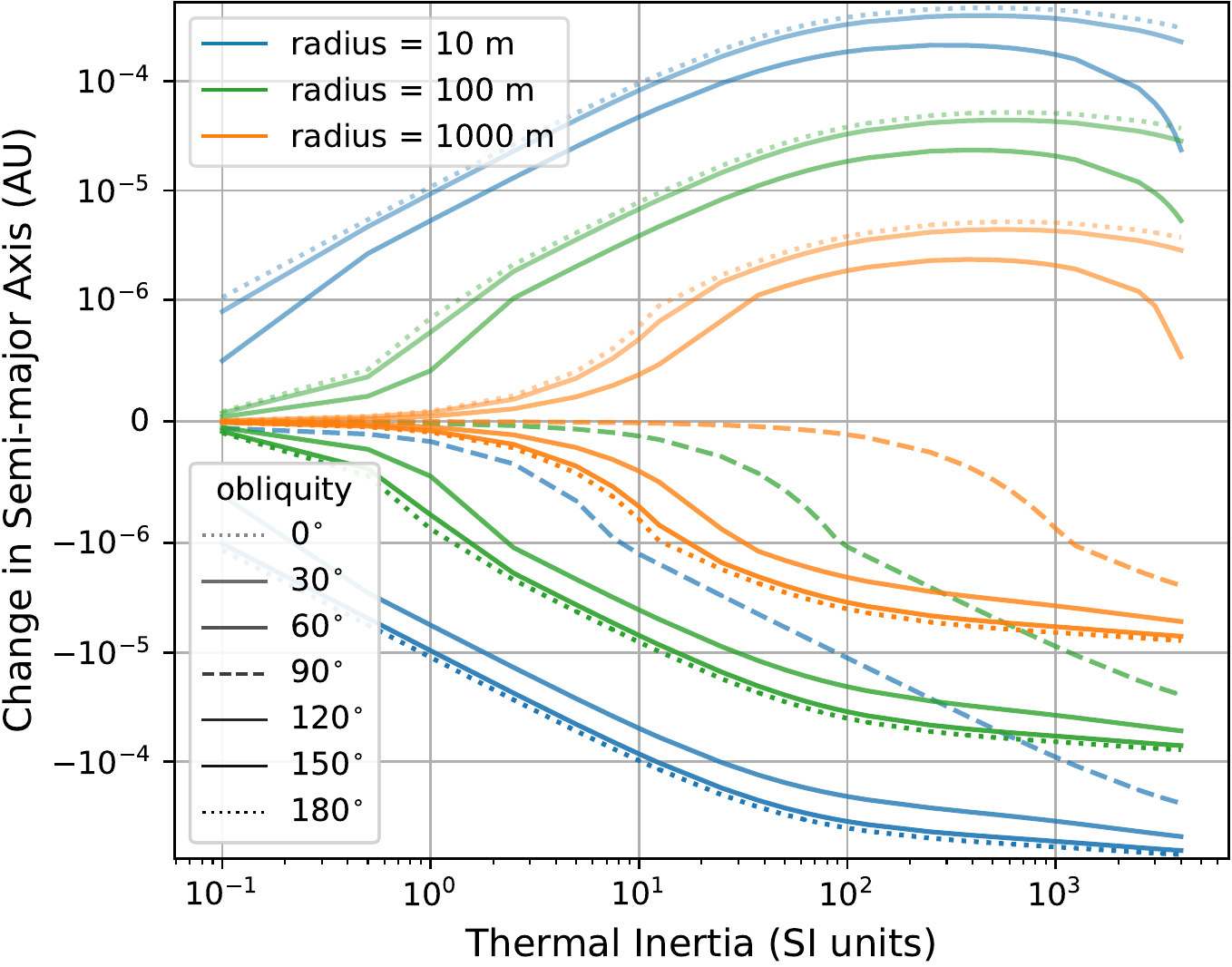}
    \caption{Change in the semi-major axis of asteroids with 
    different obliquities, thermal inertias, and
    radii after a 0.01 Myr-long simulation on a log-log plot.  Obliquities are denoted
    by opacity and radii are shown using color. A dashed line indicates an obliquity of $90^\circ$ where the body's semi-major axis is only affected by the seasonal component of the Yarkovsky effect, and a dotted line indicates an obliquity of $0^\circ$ or $180^\circ$ where the diurnal component's effect on the object's semi-major axis is at a maximum.}

    \label{fig:fig1}
\end{figure}

Figure~\ref{fig:fig1} shows how the change in semi-major axis of these simulated asteroids depends on these changing parameters.  As expected from Equation~(\ref{eq:veras2015}), the change in semi-major axis increases as the radius of the body decreases.  In addition, Equations~(\ref{eq:phi}) and~(\ref{eq:xi}) describe how bodies with small thermal inertias will radiate away too much heat before they rotate enough for the effect to activate, leading to a nearly negligible change in semi-major axis from the Yarkovsky effect. 

Figure~\ref{fig:fig1} also shows how the strength of the diurnal and seasonal components of the Yarkovsky effect depend on an object's physical properties.  For example, the lines in Figure~\ref{fig:fig1} begin to noticeably trend downwards as thermal inertia reaches values above 1000\footnote{in SI units.}.  Objects with extremely large thermal inertias will hold onto thermal energy for longer periods of time as they move throughout their orbits, leading to an increased contribution from the seasonal part of the Yarkovsky effect when compared to the diurnal portion. This, combined with the fact that the seasonal component always leads to a decrease in the body's semi-major axis, explains the increased negative change in semi-major axis for bodies with higher thermal inertias.

In addition to thermal inertia, obliquity plays a large role in determining the relative strengths of the seasonal and diurnal components of the Yarkovsky effect.  Bodies with obliquities at $0^\circ$ and $180^\circ$ will feel the strongest perturbations in their semi-major axis from the diurnal component. As the obliquity moves towards $90^\circ$, the diurnal portion's effect on the object's semi-major axis will begin to decrease and its effect on the object's inclination will start to increase.  At an obliquity of $90^\circ$, only the seasonal component will affect the body's semi-major axis. This is why objects at $90^\circ$ in Figure~\ref{fig:fig1} remain relatively unperturbed until their thermal inertias reach a high enough value where the seasonal component becomes much more significant.  In between these extreme values, the body will experience the combined effects of the diurnal and seasonal components. However, because the object's orbital mean motion is small compared to its rotational frequency, the diurnal portion will dominate for most obliquity and thermal inertia values and be the largest determining factor for the object's change in semi-major axis. This is why, in Figure~\ref{fig:fig1}, the object's overall change in semi-major axis is greater for obliquity values that are closer to $0^\circ$ or $180^\circ$ where the diurnal component is strongest in the plane of the object's orbit. Users that want to artificially decrease or eliminate the effects of the diurnal component to make the seasonal component more prominent can simply increase the body's rotation period to an arbitrarily high value in their simulations.

From this model, observational information about changes in the orbital distance of an orbiting body can be used to constrain the physical properties of that body, as was done by \citet{Vokrouhlicky2008} to constrain several physical parameters of the asteroid 1992 BF and \citet{Tardioli2017} to constrain the obliquities of multiple near-Earth Asteroids.  The Full Version of the effect can also be used to approximate the age of asteroid families by determining how the orbits of their constituent asteroids evolved since the family's creation.  This information can then be used to determine the time when these asteroids began diverging from the collision that created them. \citet{Nesvorny2004} and \citet{Carruba2017} used this technique to decrease the uncertainty in the ages of the Karin and Veritas asteroid families, respectively.

\section{Simple Version}
\label{sec:simple_version}

The Simple Version of the Yarkovsky effect in \rebx\ is based on modifications made to Equation~(\ref{eq:veras2015}) by \citet{Veras2019}.  Recall from Equation~(\ref{eq:veras2015}) that $\mathbb{Y}$ is the Yarkovsky matrix that describes an object's rotation and thermal emission.  Calculating this matrix between each time step of a \reb\ simulation is time-consuming, and using it requires many parameters.  To prevent the need for such a complicated heat conduction model, \citet{Veras2019} placed constant entries into this matrix.  While these modifications lead to a less detailed model of the Yarkovsky effect, they decrease the computational time needed for calculations.  Two of these prescriptions (models A and B) are available for use in the Simple Version of the Yarkovsky effect.

These matrices have been modified slightly to remove the effects of PR drag by changing the diagonal terms from 1 to 0.  PR drag is already a feature in the previously implemented Radiation Forces effect in \rebx\ \citep{Tamayo2020}, so its inclusion in the Simple Version would be superfluous. In addition, while PR drag is important to consider for dust-sized particles, the Yarkovsky effect plays a larger role in the orbital evolution of larger bodies.  \citet{Veras2015} showed that the Yarkovsky effect should only be active in bodies with diameters greater than 10 m that are spinning moderately fast.  In addition, \citet{Veras2015} show that the Yarkovsky effect is proportional to $1/c$ while PR drag is proportional to $1/c^2$. Therefore, when the Yarkovsky effect is active on a body, it is reasonable to ignore the effects of PR drag.  
\begin{equation}
   \mathrm{Model\ A:\ } \mathbb{Y} = \Large \begin{bmatrix} 0 & 0 & 0\\ 1 & 0 & 0\\ 0 & 0 & 0 \\ \end{bmatrix};
    \label{eq:model_A}
\end{equation}
\begin{equation}
    \mathrm{Model\ B:\ } \mathbb{Y} = \Large \begin{bmatrix} 0 & 1 & 0\\ 0 & 0 & 0\\ 0 & 0 & 0 \\ \end{bmatrix}.
    \label{eq:model_B}
\end{equation}
Model A and Model B are shown in Equations~(\ref{eq:model_A}) and~(\ref{eq:model_B}), respectively.  Model A is configured to push targets outward (corresponding to prograde rotation), while Model B is configured to drive targets inward (corresponding to retrograde rotation).  Setting the \rebx\ parameter "ye\_flag" to 1 on an object in the simulation will apply model A to $\mathbb{Y}$, while setting "ye\_flag" to -1 will apply model B to $\mathbb{Y}$.  In addition to the simplifications made to $\mathbb{Y}$, it's assumed that $k = 0.25$ for this version of the effect.  The equation for the Simple Version of the effect is then
\begin{equation}
    \left(\frac{d\textit{\textbf{v}}}{dt}\right)= \frac{3L(1-\alpha)}{64\pi \rho Rcr^2}\mathbb{Y}\textit{\textbf{i}},
    \label{eq:simple_veras2015}
\end{equation}
where $\mathbb{Y}$ will either be Model A or Model B depending on what the user chooses.  The following section will show how the change in semi-major axis of an asteroid under the influence of this version of the effect can be represented analytically while Section~\ref{sec:yark_effect_parameter_interpolation} gives a demonstration of using this version of the effect in combination with the parameter interpolation feature in \rebx.

\subsection{Analytic Comparison}
\label{sec:analytic_comparison}

If an object is in a circular orbit with no inclination around a star whose mass and luminosity are constant, then an analytic equation describing the time-evolution of the object's semi-major axis can be derived from the equations of the Simple Version. This derivation is made possible due to the Simple Version's modifications to the Yarkovsky matrix, and it cannot be performed using the equations from the Full Version. While this analytic equation only applies to the specific situation described above, we can use it to show that the Simple Version of the effect properly applies Equation~(\ref{eq:simple_veras2015}) to targets in \reb\ simulations.

Equation (A2) in \citet{Veras2015} shows how the average Yarkovsky-induced time rate of change of a body's semi-major axis depends on the entries in $\mathbb{Y}$:
\begin{equation}
    \left(\frac{da}{dt}\right)= \frac{kAL(1-\alpha)}{4\pi mnca^2(1-e^2)} \begin{bmatrix}\mathbb{Y}_{21} - \mathbb{Y}_{12} \\\ \mathbb{Y}_{32} - \mathbb{Y}_{23} \\\ \mathbb{Y}_{31} - \mathbb{Y}_{13} \\ \end{bmatrix} \cdot \begin{bmatrix} \cos(i) \\\ \sin(i)\sin(\Omega) \\\ \sin(i)\cos(\Omega)\\ \end{bmatrix},
    \label{eq:Veras_A4}
\end{equation}
where $a$ is the semi-major axis of the body, $n$ is the mean motion of the body, $i$ is the object's inclination, and $\Omega$\ is the target's longitude of ascending node.  If we assume that the object's inclination and eccentricity are both 0 and that the luminosity of its host star is constant with time, then the equation simplifies to
\begin{equation}
    \left(\frac{da}{dt}\right)= \frac{kAL(1-\alpha)}{4\pi mnca^2}(\mathbb{Y}_{21} - \mathbb{Y}_{12}).
    \label{eq:Veras_A4_with_n}
\end{equation}
To eliminate all time dependency in variables other than $a$, we must get $n$ in terms of $a$:
\begin{equation}
    n = \frac{\sqrt{G(M+m)}}{a^{3/2}},
    \label{eq:mean_motion}
\end{equation}
where $G$ is the gravitational constant and $M$ is the mass of the primary.  If we assume that the mass of the body is small compared to the mass of the primary, the body orbits a 1 $M_{\sun}$ star, and the units we use are astronomical units, $M_{\sun}$, and years so that G has a value of $4\pi^2$, then Equation~(\ref{eq:mean_motion}) simplifies to
\begin{equation}
    n = \frac{2\pi}{a^{3/2}}.
    \label{eq:mean_motion_simplified}
\end{equation}
We can now replace $n$ in Equation~(\ref{eq:Veras_A4_with_n}) with Equation~(\ref{eq:mean_motion_simplified}) to get the following separable differential equation:
\begin{equation}
    \left(\frac{da}{dt}\right)= \frac{kAL(1-\alpha)}{8\pi^2 mca^{1/2}}(\mathbb{Y}_{21} - \mathbb{Y}_{12}).
    \label{eq:Veras_A4_difeq}
\end{equation}
Solving this differential equation, we get
\begin{equation}
    \frac{2a^{3/2}}{3} = \frac{kAL(1-\alpha)t}{8\pi^2 mc}(\mathbb{Y}_{21} - \mathbb{Y}_{12}) + C,
    \label{eq:Veras_A4_difeq_solved}
\end{equation}
where $t$ is time and $C$ is a constant of integration.  We can transform Equation~(\ref{eq:Veras_A4_difeq_solved}) into a form that more resembles Equation~(\ref{eq:simple_veras2015}) by assuming $A = \pi R^{2}$ and $k$ has a value of 0.25:
\begin{equation}
    \frac{2a^{3/2}}{3} = \frac{R^2L(1-\alpha)t}{32\pi mc}(\mathbb{Y}_{21} - \mathbb{Y}_{12}) + C,
    \label{eq:Veras_A4_difeq_solved_simplified}
\end{equation}
where $R$ is the radius of the body.  By isolating $a$ and solving for $C$ when $t = 0$, the equation becomes
\begin{equation}
    a(t) = \left(\frac{3R^2L(1-\alpha)t}{64\pi mc}(\mathbb{Y}_{21} - \mathbb{Y}_{12}) + a_{0}^{3/2}\right)^{2/3},
    \label{eq:yark_analytic}
\end{equation}

\begin{figure}[htbp!]
	\includegraphics[width=\columnwidth]{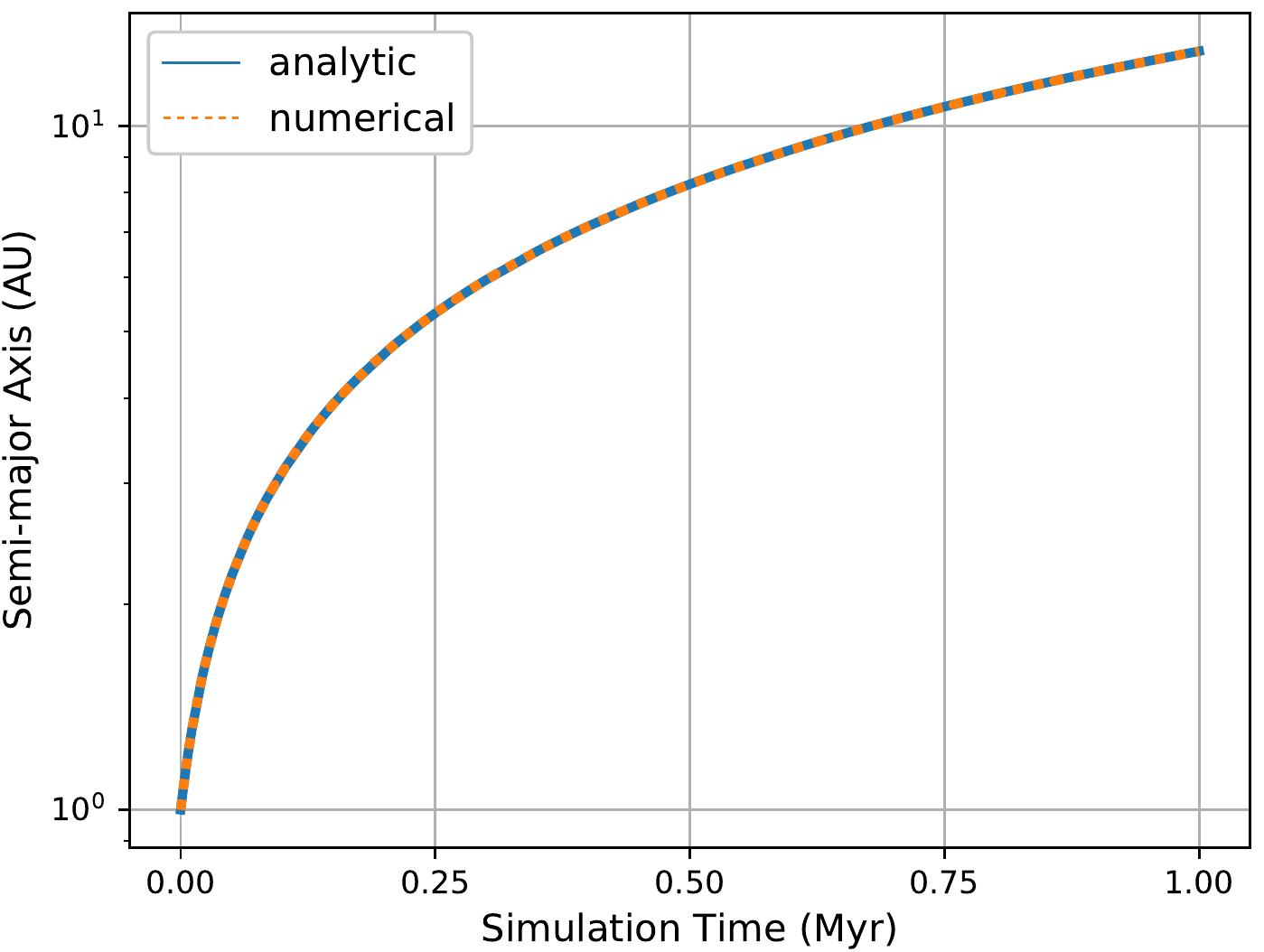}
    \caption{Change in the semi-major axis of an asteroid starting at 1 au over the course of 1 Myr.  The dashed line uses data collected from a 1 Myr simulation with the Simple Version of the Yarkovsky effect active.  The solid line is the curve created by the time evolution equation that was derived from the Simple Version.}
    \label{fig:fig2}
\end{figure}

where $a_{0}$ is the starting semi-major axis of the body.  This is the final form of the equation and can be used to replicate and validate certain results obtained from the Simple Version of the effect.  To demonstrate this, we created a simple simulation containing an asteroid in a flat, circular orbit around a 1 $M_{\sun}$ star.  The asteroid has a density of 3000 kg m$^{-3}$, a radius of 1 km, and a semi-major axis of 1 au. The star has a constant luminosity of 100,000 $L_{\sun}$ to make the effects of the Yarkovsky effect more noticeable on a shorter timescale.  For this simulation, we used Equation~(\ref{eq:model_A}), which will push the body's orbit outward.  The simulation lasts for 1 Myr and records how the semi-major axis of the body changes over time due to the Simple Version of the Yarkovsky effect.  We then compared the results from the simulation to the results when using the exact same parameters in Equation~(\ref{eq:yark_analytic}).  

Figure~\ref{fig:fig2} shows both the results from the simulation and the curve created by the analytic equation.  The two results are virtually identical for at least 1 Myr, showing that the Simple Version will properly apply Equation~(\ref{eq:simple_veras2015}) to targets in \reb\ simulations with substantial durations.  

\section{Combining Effects}
\label{sec:combining_effects}

\begin{figure*}
	\includegraphics[width=\textwidth]{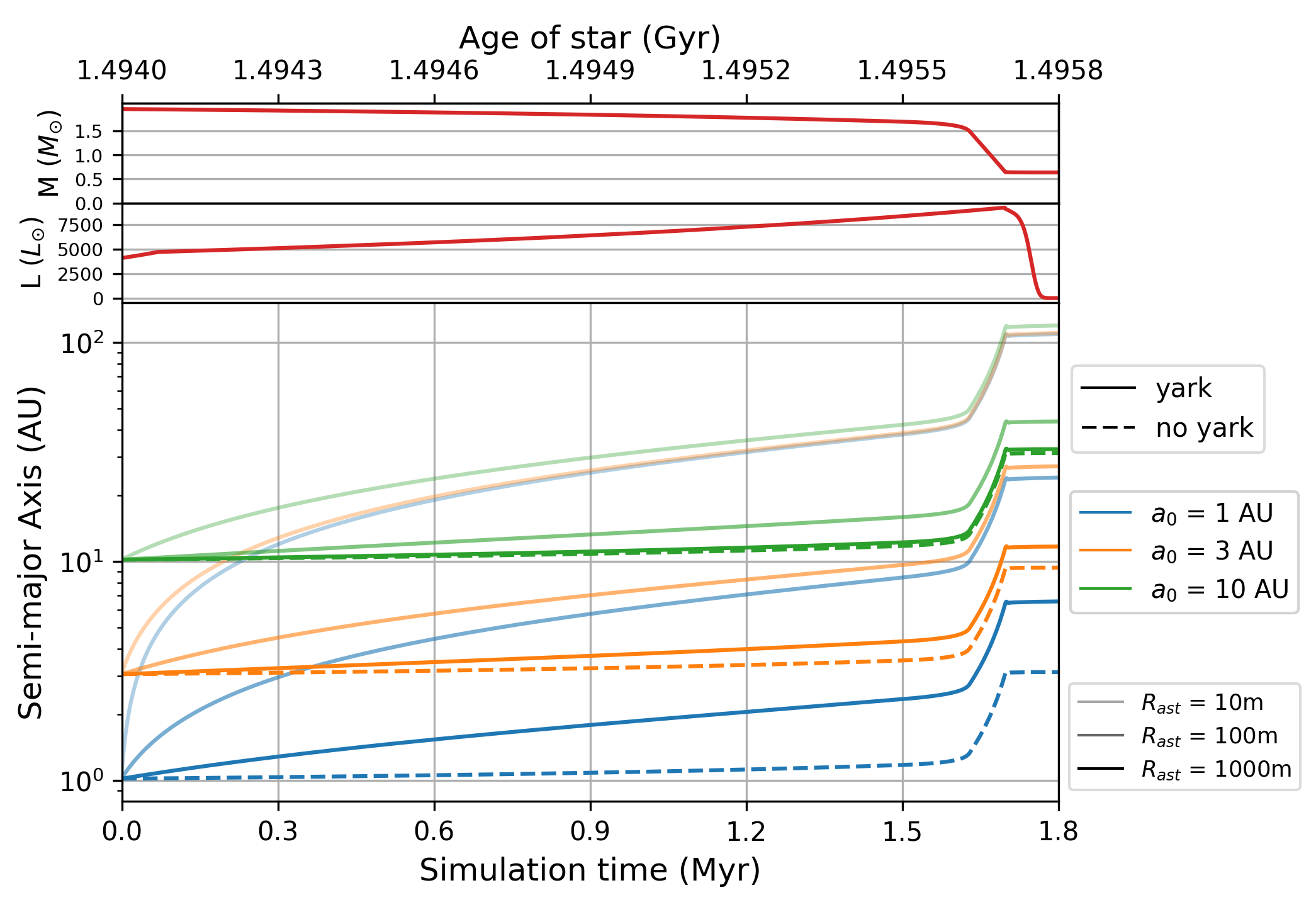}
    \caption{Simulation of asteroids with different initial conditions and physical parameters during the AGB phase of a 2 $M_{\sun}$ star.  The horizontal axis at the top of the graph shows the age of the star during the simulation, while the bottom horizontal axis gives
    the simulation time.  The top two panels show the \sse\ data for the mass and luminosity of the star during the AGB.  The bottom panel shows
    the semi-major axes of different asteroids and how they change during the length of the simulation with and without the Simple Version of the Yarkovsky effect enabled.}
    \label{fig:fig3}
\end{figure*}

\rebx\ provides users with the capability to easily use multiple forces, effects, and operators in a single simulation, allowing for detailed studies of situations that require a combination of different physics (giant branch studies, protoplanetary disks, etc.).  Section~\ref{sec:yark_effect_parameter_interpolation} shows an example that combines the Yarkovsky effect with parameter interpolation while Section~\ref{sec:time_performance} shows how the different versions of the Yarkovsky effect perform when combined with the Radiation Forces effect in \rebx.

\subsection{Yarkovsky effect and Parameter Interpolation}
\label{sec:yark_effect_parameter_interpolation}

\rebx\ effects can be used in conjunction with the parameter interpolator in \rebx, a feature that allows one to use imported time-series data for parameters that will change throughout the course of a simulation \citep{Baronett2022}.  Based on the cubic spline algorithm from \citet{Press1992}, this interpolator takes user-inputted data for a particular parameter and creates a cubic spline that can be called upon at any arbitrary time during a simulation to obtain a value for that parameter.  This interpolator is written in C, is machine-independent, and supports both forward and backwards integration. 

To demonstrate the combined capabilities of parameter interpolation and the Yarkovsky effect, we run a simple simulation based on studies done by \citet{Veras2019} of asteroids under the influence of the Yarkovsky effect during the late-stage phases of stellar evolution.  The simulation contains a star with a mass of around 2 $M_{\sun}$\footnote{The actual starting mass is 1.951859 $M_{\sun}$.} that is about to ascend the asymptotic giant branch (AGB) and asteroids placed in circular orbits at 1 au, 3 au, and 10 au.  These asteroids each have a density of 3000 kg m$^{-3}$ and can have radii of 10 m, 100 m, and 1000 m.  The Simple Version of the Yarkovsky effect is applied to these nine asteroids for the entirety of the simulation, with the effect set to propel these bodies outward.  We also add three control asteroids without the Yarkovsky effect active on them at these three different starting positions.  To save computational time and keep the focus of the simulation on the effects of stellar mass loss and the Yarkovsky effect, all asteroids in this simulation are semi-active.\footnote{Setting particles to semi-active in \reb\ simulations eliminates their gravitational influence on other bodies while retaining the gravitational influence of active particles onto them.}

The simulation lasts for 1.8 Myr and begins about 1.7 Myr before the star reaches the tip of the AGB.  The time-series data for the mass and luminosity of the originally 2 $M_{\sun}$ star were obtained from the \sse\ stellar evolution code \citep{Hurley2000}.  During the simulation, the star will lose approximately 67\% of its starting mass, which will cause all asteroid orbits to expand adiabatically by roughly a factor of three.  The luminosity of the star will reach values thousands of times greater than the luminosity of the Sun, which will cause an increase in the strength of the Yarkovsky effect for the duration of the AGB.

The top two panels of Figure~\ref{fig:fig3} show the mass and luminosity of the star during the course of the AGB phase.  The bottom panel of Figure~\ref{fig:fig3} shows the semi-major axes of the asteroids as a function of simulation time and the age of the star.  As expected, asteroids with smaller radii are propelled outward much faster by the Yarkovsky effect, and the combination of the Yarkovsky effect and stellar mass loss can push small asteroids to orbits that are over 100 times greater than their starting positions.  The three asteroids not under the influence of the Yarkovsky effect remain in relatively unchanging orbits until near the end of the simulation, when the star quickly loses a large portion of its mass.

This simulation and previous studies \citep{Zuckerman2010, Frewen2014, Veras2022} have shown that the Yarkovsky effect plays a role in determining the final arrangement of rocky material around white dwarfs and is an important mechanism for phenomena such as white dwarf pollution.  The Yarkovsky effect in \rebx\ will be a useful tool to further explore the role this radiative force has on small bodies during the late stages in a star's life.

\subsection{Time Performance}
\label{sec:time_performance}

\begin{table}
    \centering
    \caption{Time Performance of WHfast Simulations with Different Combinations of \rebx\ Effects}
    \begin{tabular}{lccc}
    \hline
    \hline
         Effects & Avg. Runtime     & Std. Dev.     & Ratio \\
                 &        (s)       &       (s)     &       \\
         \hline
         None & 0.116 & $\pm 0.010$ & ... \\
         RF & 0.153 & $\pm 0.012$ & 1.322\\
         SV & 0.154 & $\pm 0.010$ & 1.324\\
         FV & 0.327 & $\pm 0.034$ & 2.820\\
         RF \& SV & 0.212 & $\pm 0.020$ & 1.831\\
         RF \& FV & 0.357 & $\pm 0.015$ & 3.078\\
         \hline
    \end{tabular}
    \label{tab:whfast_time_table}
\end{table}

\begin{table*}
    \centering
    \caption{Average Simulation Durations for Different \rebx\ Effects and Integrators}
    \begin{tabular}{lccccc}
    \hline
    \hline
         Effects & IAS15 & JANUS & SABA & EOS & Leapfrog\\
    \hline
         None & 3.863 $\pm\ 0.221$\footnote{all entries have units of seconds.} & 0.242 $\pm\ 0.013$ & 0.381 $\pm\ 0.023$ & 0.071 $\pm\ 0.005$ & 0.035 $\pm\ 0.003$\\
         RF & 6.252 $\pm\ 0.070$ & 0.384 $\pm\ 0.020$ & 0.591 $\pm\ 0.022$ & 0.127 $\pm\ 0.020$ & 0.053 $\pm\ 0.005$\\
         SV & 6.569 $\pm\ 0.165$ & 0.673 $\pm\ 0.045$ & 0.755 $\pm\ 0.027$ & 0.174 $\pm\ 0.010$ & 0.088 $\pm\ 0.008$\\
         FV & 12.578 $\pm\ 0.093$ & 1.987 $\pm\ 0.063$ & 1.952 $\pm\ 0.051$ & 0.473 $\pm\ 0.042$ & 0.237 $\pm\ 0.011$\\
         RF \& SV & 9.158 $\pm\ 0.212$ & 0.924 $\pm\ 0.050$ & 1.058 $\pm\ 0.030$ & 0.237 $\pm\ 0.019$ & 0.118 $\pm\ 0.008$ \\
         RF \& FV & 16.466 $\pm\ 0.168$ & 2.286 $\pm\ 0.078$ & 2.255 $\pm\ 0.059$ & 0.524 $\pm\ 0.016$ & 0.287 $\pm\ 0.037$\\
    \hline
    \end{tabular}
    \label{tab:time_table}
\end{table*}

As explained in Section~\ref{sec:simple_version}, the equations contained in the Simple Version of the effect are less detailed than the ones in the Full Version to save computational time during longer simulations. To demonstrate this difference in time performance, we apply each version of the effect to a 0.01 Myr-long simulation and record its duration. The simulation contains a test particle orbiting the Sun in a circular orbit at 1 au and uses the WHfast integrator \citep{RT2015, RTB2019} with a fixed time step of 0.05 yr.  In addition to measuring the two versions of the Yarkovsky effect, we also measure the time performance of the Radiation Forces effect in \rebx, which applies both PR Drag and radiation pressure to chosen particles in a simulation. Measuring this effect will let us compare the performance of our new effect with previously implemented radiative forces.

The following configurations of effects were applied to the simulation: (1) no effects; (2) Radiation Forces; (3) the Simple Version; (4) the Full Version; (5) Radiation Forces and the Simple Version; and (6) Radiation Forces and the Full Version. We performed 10 runs for each combination and calculated their average runtimes and standard deviations.  

Table~\ref{tab:whfast_time_table} shows the results for each configuration. As expected, simulations using the Simple Version require less computational time than ones using the Full Version.  On average, we find that the Full Version increases a simulation's duration by a factor of about 2.820 while the Simple Version increases it by a factor of 1.324, which is comparable to the factor of 1.322 from Radiation Forces.  We also see a difference in computational times when Radiation Forces is combined with the two different versions of the Yarkovsky effect. Radiation Forces combined with the Simple Version increases the length of a simulation by an average factor of 1.831, which is significantly less than the factor of 3.078 from Radiation Forces combined with the Full Version. These results show that the Simple Version of the effect is more useful for long-term simulations containing many bodies, while the Full Version is more useful for short, highly-detailed simulations with fewer bodies.

For the convenience of users, we also provide time performance data for five other integrators available in \reb: IAS15 \citep{Rein2015}, JANUS \citep{Rein2018}, SABA\footnote{specifically SABA(10,6,4) (default option).} \citep{RTB2019}, EOS \citep{Rein2020}, and Leapfrog. Apart from the integrators, the setup for the simulations and methods for data collection remain the same. Table~\ref{tab:time_table} shows the average durations and standard deviations for simulations using these integrators with the same set of varied combinations of \rebx\ effects active.

\section{Conclusion}

Two different forms of the Yarkovsky effect have been added to the external library \rebx: the Full Version and the Simple Version.  The Full Version will be useful for constraining the physical properties of asteroids \citep{Vokrouhlicky2008, Tardioli2017}, estimating the ages of asteroid families \citep{Nesvorny2004, Carruba2017}, and in general, short but detailed simulations containing fewer objects.  The Simple Version will be useful for studying the evolution of smaller objects during the post-main sequence \citep{Veras2019} -- which could have important consequences for white dwarf pollution \citep{Zuckerman2010, Frewen2014, Veras2022} -- and in general, longer simulations with more particles that can tolerate lower levels of detail.  We hope these new capabilities for \rebx\ will enable new studies into the role of radiative forces in planetary systems, and we encourage others to make contributions to \rebx\ as well.

\hspace{1.5 pt}

We thank Dimitri Veras for his insight and helpful discussions and an anonymous reviewer for constructive comments that improved this article.

\textit{Software}: Simulations in this paper used the \sse, \reb, and \rebx\ codes, which are all freely available at \url{https://astronomy.swin.edu.au/~jhurley/bsedload.html}, \url{http://github.com/hannorein/rebound}, and \url{https://github.com/dtamayo/reboundx}.  This paper also made use of the open-source projects \jupyter\ \citep{Jupyter}, \texttt{IPython}
\citep{IPython}, and \texttt{matplotlib} \citep{matplotlib1, matplotlib2}.

All of the data and \python\ scripts used to generate the figures in this article are available at \url{https://github.com/Nofe4108/REBOUNDx_Paper}.


\bibliographystyle{aasjournal}
\bibliography{references}{}

\end{document}